\documentclass[11pt]{article}

\usepackage{amsmath,amssymb,amsbsy,amsfonts,latexsym,graphicx}
\usepackage{color,array,subfigure}
\usepackage{cite}
\usepackage[colorlinks,hyperindex]{hyperref} 

\newcommand{\dd}{\mathrm{d}}
\newcommand{\del}{\partial}

\allowdisplaybreaks

\evensidemargin \oddsidemargin

\begin{document}


\begin{titlepage}

\renewcommand{\thefootnote}{\fnsymbol{footnote}}


\begin{flushright}
\end{flushright}

\vspace{15mm}
\baselineskip 9mm
\begin{center}
  {\Large \bf   Character  of Matter in Holography: \\
   Spin-Orbit Interaction  
  }
\end{center}

\baselineskip 6mm
\vspace{10mm}
\begin{center}
Yunseok Seo$^1$, Keun-Young Kim$^2$, Kyung Kiu Kim$^{2,3}$  and Sang-Jin Sin$^1$
 \\[10mm]
  $^1${\sl Department of Physics, Hanyang University
    \\ Seoul 133-791, Korea}
  \\[3mm]
   $^2${\sl School of Physics and Chemistry, Gwangju Institute of Science and Technology, 
   \\  Gwangju 500-712, Korea}
   \\[3mm]
   $^3${\sl Department of Physics, College of Science, Yonsei University, Seoul 120-749, Korea}
\end{center}

\thispagestyle{empty}

\vskip 1cm

\begin{center}
{\bf Abstract}\\
\end{center}

Gauge/Gravity duality as a theory of matter needs a systematic way to 
characterise a system.  
We suggest a `dimensional lifting' of  the  least irrelevant interaction  to the bulk theory. 
As an example, we consider  the  spin-orbit interaction,  which causes   
 magneto-electric interaction term. We show that its lifting is an axionic coupling. 
We present an exact and analytic solution describing diamagnetic response. 
Experimental data on annealed graphite shows a remarkable similarity to our theoretical result.  
We also find an analytic formulas of  DC transport coefficients, according to which,  the anomalous Hall coefficient  interpolates  between the coherent metallic regime with  $\rho_{xx}^{2}$  and incoherent metallic regime with  $\rho_{xx}$  as we increase the disorder parameter $\beta$.
The  strength of the spin-orbit interaction also interpolates between the two scaling regimes. 
\vfill
\noindent
 Keywords : Holography, Diamagnetism, DC conductivity, Anomalous Hall effect. 
\\ PACS numbers : 11.25.Tq,  72.15.Eb,  

\vspace{5mm}
\noindent{\tt yseo@hanyang.ac.kr,~fortoe@gist.ac.kr,~kimkyungkiu@gmail.com,~sangjin.sin@gmail.com}

\end{titlepage}

\baselineskip 6.6mm
\renewcommand{\thefootnote}{\arabic{footnote}}
\setcounter{footnote}{0}


{\bf Overview and Summary}: 
Recently, the gauge/gravity duality \cite{Maldacena:1997re,Witten:1998qj,Gubser:1998bc} attracted much interests as a possible candidate for  a reliable method to calculate strongly correlated systems. It  is a local field theory  in one higher dimensional space  called ``bulk'', with a few classical fields coupled with anti deSitter(AdS) gravity. 
Since the strong coupling in the boundary,  
is dual to a weak coupling in the bulk, the  bulk fields  can be considered as local order parameters  of a mean field theory in the bulk. It also provided  a new mechanism for instabilities   in gravity language\cite{Gubser:2008px} which is relevant to the superconductivity \cite{Hartnoll:2008kx,denef2009landscape} and the metal insulator transition \cite{Donos:2012js}.
However, as a theory for materials, it  is still in lack of one essential ingradient, a way to distinguish one matter from the others.   Although  electron-electron interaction is traded for  
the gravity  in the bulk, we still need to specify lattice-electron interactions to characterises the system. 
  Without it, we would not know what system we are working for.  
 
 Naively one may try to introduce realistic lattice at the boundary to mimic the reality.  
 However, its effects are mostly irrelevant in the infrared(IR) limit.  
 In strong coupling limit where no quasiparticle  exists, no fermi surface(FS) exists either.  
 Actually in the absence of the FS, it is almost impossible to write down any relevant {\it interaction} term in a local field theory  in higher than 1+1 dimension.\footnote{See however ref. \cite{Faulkner:2010tq} for  semi-holographic  approach  based on IR AdS2  and its virtual $CFT_{2}$, which is different from ours.}
 Therefore non-local effect may be essential for any interesting physics in strongly interacting system. 
 One interesting aspect of a holographic theory is that any local interaction in the bulk has non-local effect in the boundary \cite{Hartnoll:2011fn}. Usually one characterise a many body system in continuum limit by a few interaction terms rather than the detail of structure.  
 Therefore,  to characterise a system in holographic theory,  what we want to suggest 
 is the dimensional-lifting, by which we mean promoting the ``system characterizing interaction'' 
 of the boundary theory  to a term in the  bulk theory  using the covariant form of the interaction. 

One may wonder  what the  gravity dual of the Maxwell theory is. 
In condensed matter, there are two components of electromagnetic interaction. One is 
electron-electron interaction and the other is lattice-electron interaction. 
While the main difficulty is coming from the former, system is characterised by the latter.  
Working hypothesis  is that the electron-electron interaction is taken care of by working in asymptotic AdS gravity. 
Our  purpose is to include the electron-lattice interaction in this holographic scheme, which is possible for two reasons.
First,   in any boundary system with a conserved  global $U(1)$ charge, we have a bulk Maxwell theory,  
which can accommodate usual electromagnetic field as a probe or an external source. 
It was used to build the holographic version of superconductivity mentioned above and also to calculate electric/thermal transport coefficients \cite{Donos:2014uba,Donos:2014cya,Kim:2015wba,Ge:2014aza}.
Second,   we can  use a relativistic theory  for a non-relativistic system.   
The relativistic invariance highly constrains the possible form of extension of interaction. 
A practical way to proceed is to turn on the interaction one by one for technically simplicity. 
The covariant form of the interaction is either scalar or top form. 
The former is trivially lifted to higher dimension, e.g,  $F_{\mu\nu}M^{\mu\nu}$ can be used in any dimension. Now suppose the top form of the boundary theory is $F_{d}$  and the bulk theory already contains scalar operator $\varphi$ and one form  $\omega_{1}$. Then we have essentially two  choices:  $\varphi \,dF_{d}$ and $\omega_{1}\wedge F_{d}$ to avoid  the total derivative term. 

To discuss the idea in more specific context, 
we consider the spin-orbit interaction  in  2+1 dimensional systems. 
It creates lots of interesting phenomena including topological insulators     
and Weyl semi-metal \cite{hasan2010colloquium,qi2011topological,fu2007topological,zhang2009topological,Landsteiner:2015pdh} by changing band structures,  
which in turn causes  magneto-electric phenomena  
\cite{li2010dynamical,kim2013dirac}  like anomalous  Hall effect.
Naively,  introducing the spin-orbit interaction involve  fermions.\footnote{The Chern-Simons term is derived from a minimal interaction ${\bar \psi}\gamma^{\mu}\psi A_{\mu}$. If we  take non-relativistic limit first,  the interaction Lagrangian is 
$L_{int}=\vec{\mu}\cdot \vec{B} $ in the electron at rest frame, which becomes ${\bar \psi}\gamma^{\mu\nu}\psi F_{\mu\nu}$   in  covariant form that is valid in any frame. When we include fermions explicitly, we have to take    into account this issue. }
However, we can integrate out the massive fermions, thereby   
 avoid dealing with fermions in our theory.  Notice that in the absence of Fermi sea as in our strong coupling problem, fermions can be considered to be  massive.  It is well known that  the fermions  integrated out leave the Chern-Simons term
 $A\wedge F$ \cite{kao1985radiatively,bernstein1985radiative}, which   can be  
 lifted to 4 dimension as   $F\wedge F \sim E\cdot B$.\footnote{Previously the Chern-Simons term in the bulk and its higher dimensional analogue were extensively 
 considered in holography  to discuss the chiral effects or instability to the inhomogeneous phases \cite{hohenegger2009note,fujita2009fractional,matsuo2010chern,donos2012spatially,nakamura2010gravity,ooguri2010holographic,domokos2007baryon,donos2011holographic,bergman2011striped} .}
 Since  it is a total derivative   by itself, we have to couple it with
  an appropriate scalar operator to have a non-trivial dynamical effect. 
In this paper, we choose it  to be the kinetic energy term of the axion scalar fields $\chi_I$. That is our interaction term is 
$\sum_{I=1,2}{\left(\partial\chi_I\right)^2}  F \wedge  F  \,,$
where $\chi_I$ was introduced to provide some disorder giving momentum dissipation \cite{Andrade:2013gsa}.
 
  Since we want to have finite temperature, chemical potential, magnetic fields, and finite DC conductivity, the system should contain metric, gauge fields and axion scalar fields ($g_{\mu\nu}, A_{\mu},\chi_{I}$) as the minimal ingredients in the bulk. So we have to start with the Einstein-Maxwell-axion system.  
We have found an exact {\it analytic} solution of such a non-trivially coupled system with a new interaction term, 
consequently yielding an explicit and analytic result for the DC conductivity using recent technology   
\cite{Donos:2014uba,Donos:2014cya,Kim:2015wba}.  
While the  Hall effect is obviously connected to our system from the construction,  the fully back reacted system shows diamagnetic response. This is because we 
examined metallic state at finite temperature and 
did not include spin degrees of freedom explicitly.   Finally,  we  comment on the relevance of our result to  experimental data. In \cite{Kopelevich:aa}, it  was   reported that graphite, once annealed to wash out the ferromagnetic behavior, shows a non-linear diamagnetic response which is very similar to  our analytic result. Also  it turns out that our analytic conductivity formulas reproduce the experimental data on the scaling relation between the non-linear anomalous Hall coefficients and the longitudinal resistiviy. i.e. the non-linear anomalous Hall coefficients interpolate between the linear and quadratic dependence on the longitudinal resistivity.  Considering that we added just one interaction term, these are  unexpectedly rich consequences. 

{\bf The model and   background solution}: 
 With motivations described above, we start from the Einstein-Maxwell-axion action with the Chern-Simons interaction
\begin{equation}\label{action0}
\begin{split}
2\kappa^{2} S = &\int d^4 x \sqrt{-g}  \left\{   R + \frac{6}{L^2}    -\frac{1}{4} F^2   -  \sum_{I=1,2} \frac{1}{2} ( \partial\chi_I  )^2      \right\} - \frac{1}{16 } \int q_\chi {\left(\partial\chi_I\right)^2}  F \wedge  F    +S_{c} \,,  
\end{split}
\end{equation}
where $q_{\chi}$ is a coupling, and  $\kappa^{2}={8\pi G}$ and $L$ is the AdS radius and we set $2 \kappa^2 = L = 1$.  $S_{c}$ is the counter term which is necessary to  make the action finite. Explicit form of $S_{c}$ is written in \eqref{action0} at the end of this paper. 
 The axion ($\chi_I$) which is linear in $\{x,y\}$ direction breaks translational symmetry and hence gives an effect of momentum dissipation \cite{Andrade:2013gsa}.    
 Instanton density  coupled with  the axion can  generate magneto-electric property: if we add charge, non-trivial magnetization is generated.   
The  equations of motion are rather long so we wrote it in \eqref{eqn1} at the end.

As ansatz to solutions, we use the following form 
\begin{equation}\label{ansatz0}
\begin{split}
&A = a(r) dt + \frac{1}{2} H\left( x dy - y dx  \right) \,,  \cr
&\chi_1 =\beta \, x \,, \quad \chi_2 = \beta \, y \,,
\end{split}
\end{equation} 
with the metric ansatz
\begin{equation}\label{ansatz}
ds^2 = - U(r)dt^2 + \frac{dr^2}{U(r)} + r^2 (dx^2 + dy^2)  \,.
\end{equation} 
From the equations of motion, we found  exact solution 
\begin{equation} \label{bgsol}
\begin{split}
&U(r) = r^2 - \frac{\beta^2}{2} - \frac{m_0}{r} + \frac{ q^2 + H^2}{4 r^2}+ \frac{ \beta ^4 H^2 q_\chi^2}{20 r^6}-\frac{ \beta ^2 H q q_\chi }{6 r^4} \,,  \cr
&a(r) = \mu -\frac{q}{r}+\frac{\beta^2 H q_{\chi}}{3 r^3} \,,
\end{split}
\end{equation}
where $\mu$ is a free parameter interpreted as the chemical potential and 
 $q$ and $m_0$ are determined by the condition $A_t (r_0) =  U(r_0) = 0$ at the black hole horizon($r_{0}$). $q$ is the  conserved $U(1)$ charge interpreted as a number density at the boundary system. 
 $m_0$ turns out to be half of the energy density \eqref{energy} and $\beta$ is related  to momentum relaxation rate.
\begin{equation}\label{m0q}
\begin{split}
&q = r_0 \mu +\frac{  1}{3}\theta H \quad \quad\hbox{with } \quad 
 \theta=\frac{\beta^2 q_{\chi}}{r_{0}^{2}} , 
\cr
&m_0 =  r_0^3+\frac{r_0^2 \mu^2 +H^2-2\beta^2 r_0^2}{4 r_0} + \frac{ \theta^{2}H^2}{45 r_0} \,.
\end{split}
\end{equation}
The solution (\ref{bgsol}) reproduces the dyonic black hole solution with momentum relaxation \cite{Kim:2015wba} when $q_{\chi}$ vanishes.  



{\bf  Diamagnetic response}:
 The thermodynamic potential density $\mathcal W$ in the boundary theory  is computed by the Euclidean   on-shell action $S^E$ of \eqref{action0}: 
$
S^E  \equiv   {\mathcal{V}_2} \mathcal W/{T} \,,   
\mathcal{V}_2 = \int \dd x \dd y
$
using the solutions \eqref{ansatz0}-\eqref{ansatz}.
\begin{align} \label{W1}
\mathcal W = - r_0^3 -\frac{1}{4  r_0}\left(\mu^2 r_0^2+ 2\beta^2 r_0^2 - 3  {H^2}\right)
+ \frac{2}{3} \mu\theta H + \frac{  7 }{45 r_0}  \theta^{2} H^{2}    .
\end{align}
The system temperature $T$ is identified with  the Hawking temperature of the black hole,
\begin{align}\label{temperature}
T=\frac{3}{4\pi } r_0 -\frac{1}{16\pi r_0^3} \left( (q-\theta H)^{2} +H^2 +2 r_0^2 \beta^2 \right) \,,
\end{align}
and the entropy density is given by the area of the horizon 
\begin{equation}  \label{entropy}
s= 4\pi r_0^2. 
\end{equation}
We have numerically checked that the entropy is a monotonically increasing function of temperature for the parameters analysed in this paper.
The energy density $\varepsilon$ is one point function of the boundary energy momentum tensor $T_{00}$, which is holographically encoded in the metric \eqref{m0q}:
\begin{align}\label{energy} 
\varepsilon = \langle T_{00} \rangle = 2 m_0 .
\end{align}

It is remarkable that the complicated expression of the thermodynamic potential density \eqref{W1} gives a simple  thermodynamic relation
\begin{align} \label{W11}
\mathcal W= \varepsilon -s T - \mu q \,,
 \end{align}
 with  energy, temperature and entropy given by   \eqref{temperature}, \eqref{entropy} and  \eqref{energy}.
The variation of the potential density \eqref{W1} boils down to 
\begin{align}\label{dW2}
\delta {\mathcal W} = - \tilde{M}  \delta H - s \delta T -q \delta \mu \,,
\end{align}
where
\begin{align}
\tilde{M} &\equiv \frac1{r_0} \left(- {H} + \frac{1}{3  } \theta q -\frac{ 1}{5}\theta^2 H \right) \,, 
\end{align}
Notice that the (\ref{W11}) and (\ref{dW2}) implies the first law of thermodynamics;
\begin{align}\label{delE}
\delta \varepsilon =-\tilde{M} \delta H + T \delta s + \mu \delta q .  
\end{align}
Two important remarks are in order: 
First, $H$ is interpreted as an externally applied field, although it is a fully back reacted object in the bulk.  $H$ is the magnetic  field generated by  free current, not the magnetic induction which is usually denoted by $B$. This is because we did not encode any spin dynamics in the bulk and we do not have  fully dynamical gauge  fields at the boundary. The Maxwell fields at the boundary  enter as an external source or as a weak probe field.  

Second,  $\tilde{M}$  has dimension 1 and  describes a genuine 2+1 dimensional   system. Therefore, 
it can not be identified as the magnetisation $M$ of a physical system  which is a  2 dimensional array in 3 spatial  dimension. Furthermore, since $\tilde M$ and $H$ are different in mass dimention, they  cannot be added to form magnetic induction $B$. 
  The magnetic field $H$ and  magnetization $M$ are those of spatial 3 dimension, therefore both  $H$ and $M$ should have the same mass  dimension 2. 
If we just multiply $\tilde M$ by  $r_{0}$, a mass dimension 1 parameter which is a constant for adiabatic processes,  we would get  $B=H+M=0$ for $q_{\chi}=0$. It means that  the dyonic  black hole exhibits the Meisner effect, which is not physical. 
The problem can be traced to the fact that  after we scale $\varepsilon$ by $r_0$ to balance the dimension of $M$ and $H$, 
 the free energy contains $H^2/2$, which is the field energy of magnetic field applied on vacuum. 
  When we calculate the magnetization by taking its derivative, we should   subtract it from the free energy as suggested  by Landau and Lifshitz in section 32 of ref. \cite{landau1984electrodynamics}.   
Therefore, we   calculate the magnetization from  $F \equiv  r_0 \varepsilon -\frac{1}{2}H^2$:
 \begin{align}\label{Magnetization}
 M =-\frac{\partial F}{\partial H} \Big|_{\mathrm{fixed} ~ r_0, q }  
 =  \theta q /3 - \theta^2 H/5. 
 \end{align} 
Both terms  here are the consequences of the axionic coupling. 

The first term is the magnetization at $H=0$, which will be denoted by $M_{0}$. 
It is proportional to the charge of the system and gives ferromagnetism. More explicitly, 
\begin{align}
\label{MzeroB} 
M_{0} = \frac{1}{3} \theta q = \frac{ \mu \beta^2 q_{\chi} }{3r_{0}}, 
\end{align}
with $r_{0} =  ({  4\pi T +\sqrt{16 \pi^2 T^2 +3 (\mu^2 + 2 \beta^2)}})/6$ at ${H=0}$.
For given   $\mu, \beta$ and $q_{\chi}$, $M_0$ has the maximum value at zero temperature
and decreases as $1/T$ for large  temperature. 
In the coherent metallic regime \cite{Kim:2014bza} $ \beta/\mu \ll 1$,   $M_0 \sim  { \beta^2 q_{\chi}}  $. 
The second term in \eqref{Magnetization}  represents the back reaction of the system to the external magnetic field and gives diamagnetism.  

\begin{figure}[]
\centering
    \subfigure[]
   {\includegraphics[width=7cm]{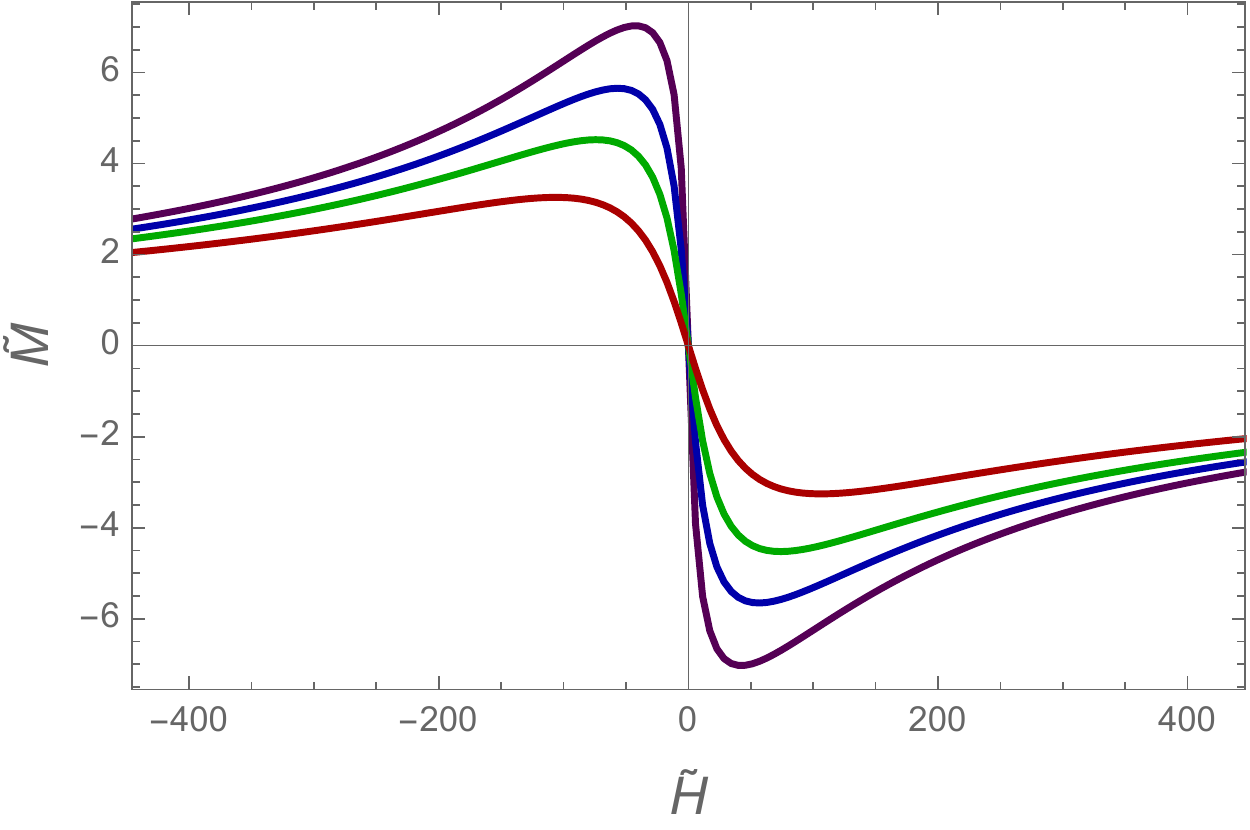} \label{}}
   \hspace{1cm}
       \subfigure[]
   {\includegraphics[width=7.7cm]{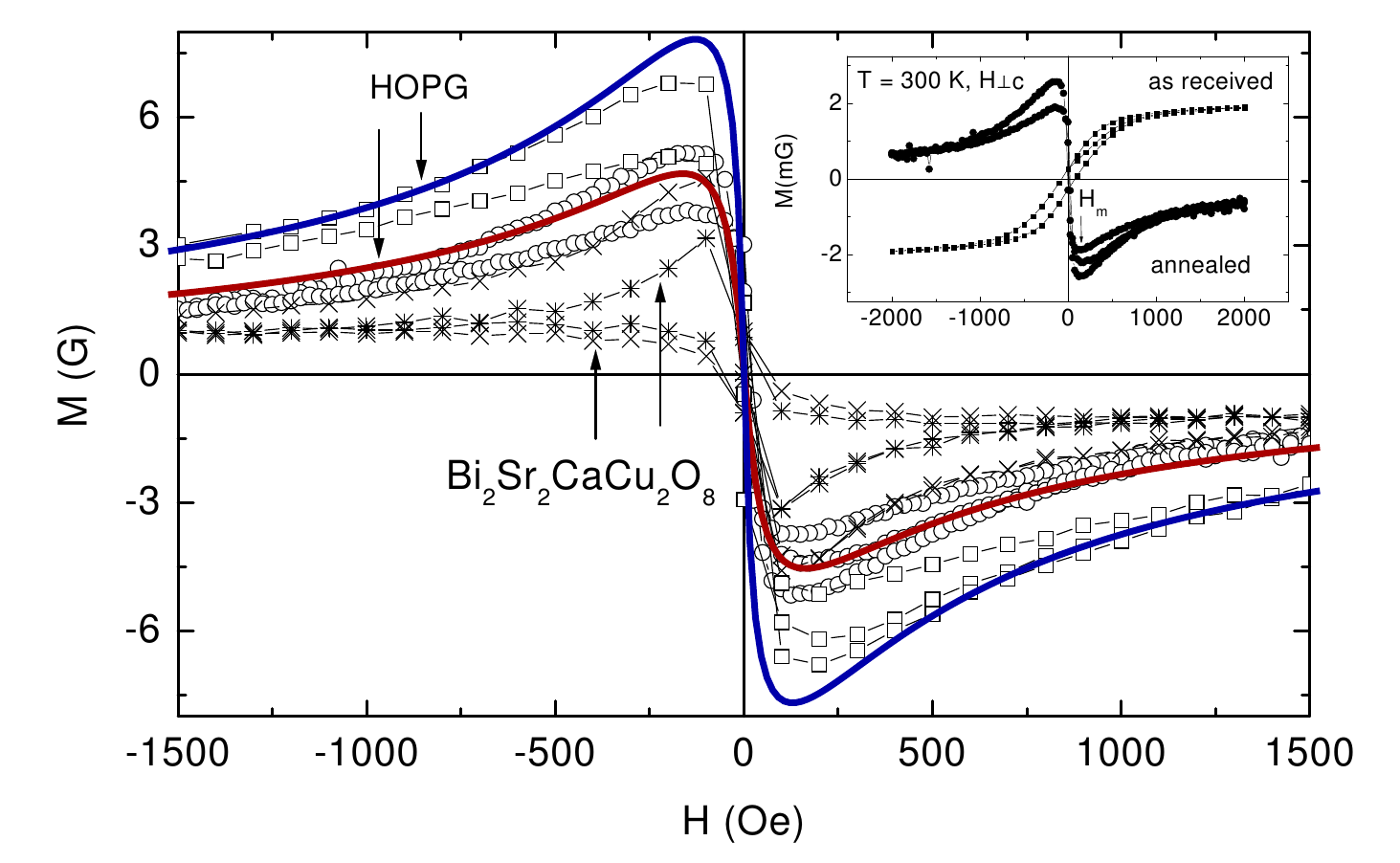} \label{}}
 \caption{(a) The external magnetic field dependences of the magnetization at $T=0.8$(purple), $T=1$(blue), $T=1.2$(green)  and $T=1.5$(red) with $q_{\chi}=10$, $\beta=2.2$, $\mu=0$. (b) Theory curve compared with experimental data of annealed highly oriented pyrolytic graphite(HOPG) sample for $T=150K$(rectangle) and $T=300K$(circle) in \cite{Kopelevich:aa}.  For comparison, the blue and red curves obtained by our formula  \eqref{Magnetization} are added. The red one is for $T=1$, $\beta=2$ and the blue one is for $T=0.8$, $\beta=2.2$. $q_{\chi}=10$ and $\mu=0$ for both cases. 
           } \label{fig:HMfixedT}
\end{figure}
We want to analyse the magnetisation  as a function of the magnetic field with the other parameters fixed.
Notice that, at fix temperature, $r_0$ has to be computed from (\ref{temperature}). 
In Figure \ref{fig:HMfixedT} (a), we draw the magnetic field dependence of the magnetization at different  temperatures for $\mu=0$, $q_{\chi}=10$, and $\beta=2$. The magnetization seems to be saturated  for large magnetic field and the magnetic susceptibility  is decreasing function of  temperature.  Our results are very similar to the graphite data in Figure \ref{fig:HMfixedT} (b) \cite{Kopelevich:aa}. Here, in addition to experimental data, we added the blue and red curves using our formula  \eqref{Magnetization} for comparison, where the red one is for $T=1$, $\beta=2$ and the blue one is for $T=0.8$, $\beta=2.2$. $q_{\chi}=10$ and $\mu=0$ for both cases.   

{\bf DC transport coefficients}: Recently, a systematic way to compute the DC transport coefficients  has been developed in \cite{Donos:2014cya, Kim:2015wba, Blake:2015ina} by which we can compute the longitudinal and transverse electric and thermoelectric conductivities. Here we write only result. 
\begin{equation}\label{dcS}
\begin{split}
\sigma_{xx} &=\frac{({\cal F} -  H^2)({\cal F} +{\cal G}^2)}{({\cal F}^2 + H^2 {\cal G}^2 )} \,,  \cr
\sigma_{xy} & =\frac{  H {\cal G} (2{\cal F}+{\cal G}^2-H^{2})+\theta ({\cal F}^2+ H^2 {\cal G}^2) }{ ({\cal F}^2 +  H^2 {\cal G}^2)} \,, \cr 
\alpha_{xx} &= \frac{  s\, {\cal G}({\cal F}-  H^2)}{{\cal F}^2 +  H^2 {\cal G}^2}\,,   \qquad
\alpha_{xy} = \frac{  s  \, H ({\cal F}+ {\cal G}^2)}{{\cal F}^2 +  H^2 {\cal G}^2}\,,
\end{split}
\end{equation}
where
\begin{equation}\label{FG}
\begin{split}
{\cal F} &= r_0^2 \beta^2 +H^2 \left( 1+ \theta^2 \right) - \theta q H  \,,  \cr
{\cal G} &= q -\theta H , \qquad s=4\pi r_{0}^{2}  \,.
\end{split}
\end{equation}
We gave some details  at the end. 
We check two limits: i) for $\beta=0$, the DC conductivities \eqref{dcS}  become $ \sigma_{xx} =0$, $\sigma_{xy} = q /H$, $\alpha_{xx}=0$ and $\alpha_{xy}=s/H$, 
which agrees with \cite{Hartnoll:2007ai};   
ii) for $q_{\chi} =0$ \eqref{dcS}  reproduces the result obtained  in \cite{ Kim:2015wba, Blake:2015ina}. 
 
At finite $\beta$ and finite $q_\chi$ but with $H=0$ the electric conductivities reduce to :
\begin{equation}\label{dcS2}
\sigma_{xx} = 1+\frac{\mu^2}{\beta^2} \,, \qquad 
\sigma_{xy} =  \theta = \frac{3 M_0}{q} .
\end{equation}
Notice that $\sigma_{xx}$ is a known result \cite{Andrade:2013gsa},  but, interestingly, $\sigma_{xy}$ is non zero even when $H=0$.   This phenomena is related to {\it anomalous} Hall effect, which will be discussed next. It is the result of the axion coupling we introduced, which gives a ferromagnetism with the magnetization $M_0$ \eqref{MzeroB}. 

{\bf Anomalous Hall effect}: 
 In ferromagnetic conductor, 
 the Hall effect is about 10 times bigger than in non-magnetic material. This stronger Hall effect in ferromagnetic conductor is known as the anomalous Hall (AH) effect \cite{Nagaosa:2010aa}.   
The  precise mechanism for AH effect has a century-long history of debates \cite{Nagaosa:2010aa}. Three mechanisms have been suggested : i) intrinsic one due to anomalous velocity, 
ii) side jump,  iii) skew scattering.   Mechanism i) was suggested   in 1950's by Karpulus and Luttinger. In modern days, the anomalous velocity is understood by the Berry phase (${\bf v}_{a}=\frac{e}{\hbar}{\bf E \times b}_{\rm Berry}$). Side jump mechanism is suggested by Berger in ref. \cite{berger1970side} where 
he showed that the electron velocity is deflected in opposite directions by the opposite electric fields experienced upon approaching and leaving an impurity.
The skew scattering  was suggested by Smit in \cite{smit1955spontaneous,smit1958spontaneous} where  
he noticed that  asymmetric scattering from impurities is caused by the spin-orbit interaction. 
The fundamental interaction underlying all these three is the spin-orbit interaction. 

It has been known that there is a power law relationship between the anomalous part($R_S$) of the Hall  resistivity($\rho_{yx}$)  and the longitudinal resistivity($\rho_{xx}$):
\begin{align}
R_s\sim \rho_{xx}^{\alpha} \,,
\end{align}
with the anomalous Hall coefficient, $R_S$, defined by the relation
\begin{align}\label{Ryx}
 \rho_{yx} = R_{H} H + R_{S} M,
\end{align} 
where $R_H$ is the usual Hall coefficient. The power $\alpha$ had been computed for  three scenarios to give $\alpha=2$ for i), ii) and  $\alpha=1$ for iii). 
 
From \eqref{Magnetization} and \eqref{dcS2} our model describes a ferromagnetic conductor, therefore it will be interesting to study AH effect. 
The resistivity matrix($\rho$) can be computed by inverting the  conductivity matrix($\sigma$) in \eqref{dcS} i.e.
${\cal \rho} = {\cal \sigma}^{-1}$.
$R_S$ is identified by  $\rho_{yx}$ at $H=0$: 
\begin{align}
\rho_{yx} \Big|_{H=0}= \frac{\theta}{ \theta^2 + (1 +\mu^{2}/\beta^2)^{2}} =R_{S}M_0.
\end{align}
Since $M_{0}=\frac13 q\theta$, anomalous Hall coefficient is given by 
\begin{align}
R_S =\frac{3 }{r_{0}\mu}\frac1{ \theta^2 + (1 +\mu^{2}/\beta^2)^{2}} .
\end{align}
The longitudinal resistivity at $H=0$  reads
\begin{align}
\rho_{xx}=\frac{ 1 +\mu^{2}/\beta^2}{\theta^2 + (1 +\mu^{2}/\beta^2)^{2}} \, .
\end{align}
 
The scaling behaviours can be read easily for   two   limits: 
\begin{align} 
R_S\sim \rho_{xx}^{2}, \quad \hbox{or} \quad R_S \sim  \rho_{xx}, 
 \end{align}
depending on $\beta/\mu \ll 1$  or  $\beta/\mu \gg 1$.
 The same scaling relations hold for    $\theta \ll 1$  or   $\theta \gg 1$. 
\begin{figure}[]
\centering
  {\includegraphics[width=8cm]{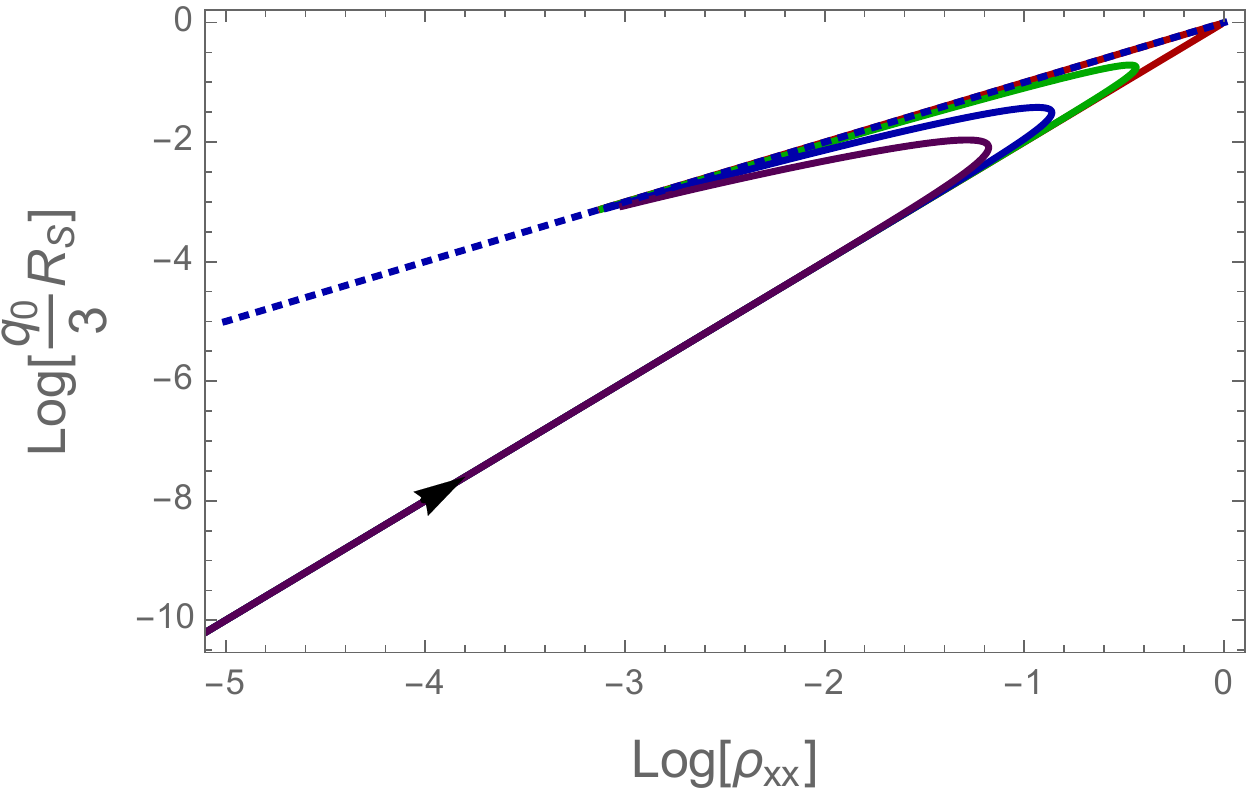} \label{}}
 \caption{ Relation between $\rho_{xx}$ and $ \frac{q_0}{3}R_{s}$($q_0 \equiv \mu r_0$) at fixed $q_{\chi}=1$ with $\mu/T =0.1$(red), $2$(green), $5$(blue) and $10$(purple) in log-log plot. The arrow represents the direction of increasing $\beta/\mu$.   $R_{S} \sim \rho_{xx}^2$  in small  $\beta/\mu$ and $R_{S} \sim    \rho_{xx} $ in large  $\beta/\mu$ regime.      } \label{fig:RsRx}
\end{figure}
%
For general value of $\beta/\mu$, the scaling behaviour  is shown in Figure \ref{fig:RsRx} as the slope. 
In the figure, 
the arrow represents the directions of increasing  disorder parameter $\beta$, and 
$\mu/T =0.1$(red), $2$(green), $5$(blue) and $10$(purple). 
Notice the  transition from $\alpha=2$  to  $\alpha=1$  is sharper for larger temperature. 
Notice  also that  $\alpha=2$ for small $\beta/\mu$ whatever is the spin-orbit interaction strength $\theta$.   

In ref. \cite{Kim:2014bza}, material  with  $\beta/\mu \ll 1$  was identified as a coherent metal of which optical conductivity has a well defined Drude peak  as if it had quasi particles. 
For  $\beta/\mu \gg 1$, the system behaves as an incoherent metal without a Drude peak. 
Interestingly, such electrically classified coherent/incoherent metal also shows AH property with the characteristic power $\alpha = 2,1$. 
This is consistent with the interpretation of $\beta$ as impurity density since large $\beta$ would have much extrinsic disorder effects.  
 Notice that  we have   two scaling regimes    in one model with interpolating parameters given by the disorder parameter $\beta$ or spin-orbit coupling strength $\theta$, while in  conventional method different scalings are associated with different mechanisms.  The fact that all three mechanisms are originated from spin-orbit interaction is reflected to our result which is a consequence of  adding just one  interaction term representing spin-orbit coupling.  
  


\begin{appendix}
{\bf Method: DC conductivites from black hole horizon}\\
Here we explain how to obtain the DC conductivities (\ref{dcS}). 
The action with counter terms are given by 
\begin{equation}\label{action0}
\begin{split}
2\kappa^{2} S = &\int_{\mathcal M} d^4 x \sqrt{-g}  \left\{   R + \frac{6}{L^2}    -\frac{1}{4} F^2   -  \sum_{I=1,2} \frac{1}{2} ( \partial\chi_I  )^2      \right\} - \frac{1}{16 } \int_{\mathcal M} q_\chi {\left(\partial\chi_I\right)^2}  F \wedge  F    +S_{c} \,, \cr 
& S_{c}= -  \int_{\partial  \mathcal{M} }d^3 x  \sqrt{-\gamma}  \left( 2 K +  \frac{4}{L} + R[\gamma] -\sum_{I=1,2} \frac{L}{2}  \nabla \chi_I \cdot \nabla \chi_I    \right) \,.
\end{split}
\end{equation}
The  equations of motion are   
\begin{equation} \label{eqn1}
\begin{split}
&\nabla^2 \chi_I +\frac{q_\chi }{8} \nabla_M \left(  \frac{1}{\sqrt{-g}} \epsilon^{PQRS} F_{PQ}F_{RS} \nabla^M \chi_I  \right)= 0 \,,  \cr
&\nabla_M F^{MN} + \frac{  q_{\chi}}{4} \nabla_M  \left(     ( \partial \chi_I )^2   \frac{1}{ \sqrt{-g}}  \epsilon^{MNPQ} F_{PQ}  \right)=0 \,,  \cr
&R_{MN} - \frac{1}{2} g_{MN} \left(   R +6 - \frac{1}{4} F^2 - \frac{1}{2} ( \partial \chi_I )^2 \right) - \frac{1}{2} F_{MP} {F_N}^P  - \frac{1}{2} \partial_M \chi_I \partial_N \chi_I  \,, \cr
&\qquad \qquad -\frac{ q_\chi}{16} \partial_M \chi_I \partial_N \chi_I \frac{1}{\sqrt{-g}} \epsilon^{PQRS}F_{PQ}F_{RS}=0 \,,
\end{split}
\end{equation}
where  $\epsilon^{0123}=\epsilon^{txyr}=1$.  

Once we get a background solution of these equations, we can compute the DC transport coefficients from the black hole horizon data. Let us start by defining a useful quantity:  
\begin{align}
{\mathcal F}^{MN} \equiv \sqrt{-g} F^{MN} + \,\frac{q_{\chi}}{4}\left(\partial \chi\right)^2 \epsilon^{MNPQ} F_{PQ}~~.
\end{align}
Then the Maxwell equation and the boundary current can be written as
\begin{align}
\partial_M {\mathcal F}^{MN} = 0~~,~~J_{tot}^{\mu} = \lim_{r\to\infty} {\mathcal F}^{\mu r} \,.
\end{align}
If we assume all fields depend on $t$ and $r$, one can obtain another expression for the total boundary current using the Maxwell equation.
\begin{align}\label{total current}
J_{tot}^\mu = \lim_{r\to r_0} {\mathcal F}^{\mu r} + \int_{r_0}^{\infty} dr  \partial_t {\mathcal F}^{t\mu}~~.
\end{align}

Now we want to consider fluctuations corresponding to the boundary DC electric field $E_i$ and the DC temperature gradient $\zeta_i=\frac{\nabla_i T}{T}$. Following \cite{Donos:2014cya},  we may consider fluctuations around the background as follows:  
\begin{equation}\label{fluctuation ansatz}
\begin{split}
\delta A_{i} &= -\left(E_i -\zeta_i a(r) \right)\, t  +\delta a_i (r)  \,, \cr
\delta g_{ti} &= -U(r) \,\zeta_i \,t + r^2 \delta h_{ti} \,, \cr
\delta g_{ri} &=r^2 \delta h_{ri} \,, \cr
\delta \chi_i &=  \delta \chi_i(r)~~\,.
\end{split}
\end{equation}
In the linear level  the time dependent part of the equations of motion drops out by the above choice of fluctuations. Thus these fluctuations are stable static fluctuations to the DC sources, $(E_i, \zeta_i)$.  

To be a physical fluctuation in the black hole background, the fluctuation should satisfy the in-falling boundary condition as it approaches the horizon. This condition can be described in terms of the Eddington-Finkelstein coordinates $v= t + \frac{1}{4\pi T}\ln \left(r-r_0\right)$, and the in-falling fluctuation should depend on $v$ near horizon.  Therefore, the regularity at the horizon implies
\begin{equation}\label{horizon}
\begin{split}
\delta h_{ti} &\sim  -  \frac{\zeta_i }{4\pi T}    \log  \left( r - r_0 \right)  \frac{U(r)}{r^2} +   h_{ti}^{(0)} + {\cal O}((r-r_0)) \,, \cr
\delta h_{ri} & \sim \frac{\mathbb H_{ri} }{r^2 U(r)}  + h_{ri}^{(0)} + {\cal O}((r-r_0)) \,, \cr
\delta a_{i} & \sim -\frac{E_i- \zeta_i a(r)}{4\pi T} \log (r-r_0) +a_i^{(0)}+ {\cal O}((r-r_0))  \,, \cr
\delta \chi_{i} &\sim \delta \chi_i^{(0)} +{\cal O}((r-r_0)) \,.
\end{split}
\end{equation}
By expanding $(r,i)$ component and $(t,i)$ component  of the Einstein equations, it turns out that $\mathbb H_{ri} =r_0^2 h_{ti}^{(0)}$ and $h_{ti}^{(0)}$ can be determined\footnote{Since the expression is lengthy but not very illuminating  we don't present here.}.

Now we are ready to consider the total current (\ref{total current}) in the linear level. The current can be calculated by plugging  (\ref{bgsol}) and the fluctuation (\ref{fluctuation ansatz}). Firstly, the second part is 
\begin{align}
\int_{r_0}^{\infty} dr  \partial_t {\mathcal F}^{t i}= \frac{\epsilon_{ij}}{r_0} \left(- {H} + \frac{1}{3  } \theta q -\frac{ 1}{5}\theta^2 H \right)\zeta_j \,.
\end{align}
This is nothing but the magnetization current
 $J^i_{mag} = - \epsilon^{ij} \tilde{M}\frac{\del_j T}{T}$
\cite{Cooper:1997aa}  which should be subtracted. 
Thus the relevant part of the current is the first term, which can be written in terms of the horizon data:
\begin{align}
J^\mu=\lim_{r\to r_0} {\mathcal F}^{\mu r} = E_i -\left( q -\theta H\right) h_{ti}^{(0)} - \frac{H}{r_0^2} \epsilon_{ij} \mathbb H_{rj}+ \epsilon_{ij} \theta E_{j}~~.
\end{align} 
Finally, the expressions $h^{(0)}_{ti}$ and $\mathbb H_{ri}$ obtained from the Einstein equation give us the electric conductivities and the thermoelectric coefficients (\ref{dcS}) using formula $\sigma_{ij} = \frac{\partial J^i}{\partial E_j}$ and $\alpha_{ij} = \frac{1}{T} \frac{\partial J^i}{\partial \zeta_j}$.

\end{appendix}

{\bf Summary and Discussion}: 
We view the holographic principle as  a set of axioms to calculate strongly interacting systems. 
For  reader's convenience we first list them below \cite{Hartnoll:2011fn}.
\begin{enumerate}
\item For  a strongly interacting system with conformal symmetry at UV, there is a dual gravity with asymptotic AdS boundary. Non AdS geometry may be regarded as an IR part of an asymptotic AdS geometry. 
\item  To calculate the correlation function of an operator ${\cal O}_{\Delta}$ with dimension $\Delta$ and spin $p$, we introduce a source field $\phi_{0}(x)$ with spin $p$ and dimension $\Delta$. 
Extend $\phi_{0}(x)$ into one higher dimensional space   $\phi(x,r)$  such that $\phi(x,r=\infty)\sim \phi_{0}(x)/r^{d-2p-\Delta}$. 
Identify the generating function of conformal field theory $Z[\phi_{0}]$   with that of  gravitational system $exp(-S[\phi_{0}])$.   
\item For a global symmetry at the bulk, we have a local gauge symmetry at the boundary. 
\item For Euclidean Green functions, Dirichlet boundary value at infinity is enough. 
For causal green function, assign  the boundary condition (BC) at the IR region   
in addition to the Dirichlet BC at the infinity. 
\item Temperature and chemical potential are provided by regularity of metric, gauge fields at the horizon or its replacing IR geometry. 
\item 
Characterise the  system by lifting the least irrelevant interactions at the boundary to the bulk. For strongly correlated electron system, the electron-electron interaction is counted by the gravity, but electron-lattice interactions should be taken into account explicitly. The gauge field dual to the conserved $U(1)$ current can have interaction terms to take care of electron-lattice interactions. 
\end{enumerate}
The last item is what we added in this paper. 
In this paper, we considered the magneto-electric phenomena induced by the spin-orbit coupling interaction as an example of dimensional  lifting  in holographic theory. 
When electron spins are correlated,  adding charge carrier changes the magnetic property as well as the charge transport.  In effective field theory approach,  such   electric-magnetic  effect  can be implied  
by adding the Chern-Simons term $\sim A\wedge F$ in  2+1 dimension. It act  as a crossing source of electricity and magnetism. 
With such a term, the system can pick up  a magnetic source when we provide electric charge and vice versa. 

We  work at  finite temperature, chemical potential, and magnetic fields. The metric, gauge  and axion  fields ($g_{\mu\nu}, A_{\mu},\chi_{I}$) are playing the role of  coupled order parameters. 
 We have found the exact and analytic solution of such a complicated coupled system with a non-trivial interaction, which made it possible to get an explicit and analytic  DC conductivity formulas. 
     
     Our results on the Hall resistivity  shows a non-linear diamagnetic response is similar to  that of graphite system and also to high $T_c$ superconductor $\rm Bi_{2}Sr_{2}CaCu_{2}O_{8}$.  Also we have shown that the anomalous Hall coefficients in our model interpolate between the linear and quadratic regime on the resistivity dependence as a function of disorder parameter and spin-orbit interaction coupling.  
It is particularly interesting to see that electrical coherent/incoherent metal has  magnetic behaviour with 
quadratic/linear resistivity dependence. 
Experimentally diverse materials were studied.  Some shows $\alpha$ near 2 and the other 1,  
sometimes old and new data  crashes. So, Detailed data mining is postponed to future investigation.

Our model does not include paramagnetic behaviour because we integrated out fermion and therefore spin degrees of freedom is not included explicitly. Also we could have chosen the scalar field $\chi$ itself instead of its kinetic term as a scalar partner of $F\wedge F$ term.  
We will  report on these issues elsewhere. 

\newpage

\section*{Acknowledgments}
The authors want to thank Junghoon Han for suggesting to look graphite data for diamagnetism. 
SS appreciates discussions with Yunkyu Bang, Yongbaek Kim, Kwon Park and especially Kiseok Kim 
on  many related  issues. YS want to thank KIAS for the financial support during his visit. 
The work of SS and YS was supported by Mid-career Researcher Program through the National Research Foundation of Korea (NRF) grant No. NRF-2013R1A2A2A05004846. The work of KKY and KKK was supported by Basic Science Research Program through the National Research Foundation of Korea(NRF) funded by the Ministry of Science, ICT \& Future Planning(NRF-2014R1A1A1003220)  and the 2015 GIST Grant for the FARE Project (Further Advancement of Research and Education at GIST College). 
\bibliographystyle{unsrt}
\bibliography{YSeoRefs}

\end{document}